\documentclass[aps,twocolumn,pra,showpacs,floatfix,groupedaddress]{revtex4-1}

\usepackage{amssymb}
\usepackage{amsmath}
\usepackage{graphicx}
\usepackage{comment}
\usepackage{color}
\newcommand{\tm}{\tablenotemark}

\usepackage{hyperref} 
\usepackage{xcolor}
\definecolor{cite}{rgb}{0.,0.,0.85}   
\hypersetup{colorlinks,linkcolor={cite},citecolor={cite},urlcolor={cite}}

\newcommand{\bra}[1]{\ensuremath{\langle #1|}}	
\newcommand{\ket}[1]{\ensuremath{|#1\rangle}}	

\usepackage[normalem]{ulem}
\definecolor{newc}{rgb}{0.,0.6,0.4}

\begin{document}
\title{Correlation trends in the hyperfine structure for Rb, Cs, Fr and high-accuracy predictions for hyperfine constants}
\author{S. J. Grunefeld}
\author{B. M. Roberts}
\author{J. S. M. Ginges}
\affiliation{School of Mathematics and Physics, The University of
  Queensland, Brisbane QLD 4072, Australia}

\date{\today}

\begin{abstract}
We have performed high-precision calculations of the hyperfine structure for $n\, ^2S_{1/2}$ and $n\, ^2P_{1/2}$ states of the alkali-metal atoms Rb, Cs, and Fr across principal quantum number $n$, and studied the trend in the size of the correlations. 
Our calculations were performed in the all-orders correlation potential method. 
We demonstrate that the relative correlation corrections fall off quickly with $n$ and tend towards constant and non-zero values for highly-excited states.
This trend is supported by experiment, and we utilize the smooth dependence on $n$ to make high-accuracy predictions of the hyperfine constants, with uncertainties to within 0.1\% for most states of Rb and Cs.
\end{abstract}


\maketitle

\section{Introduction}

The hyperfine structure lies at the interface of atomic and nuclear physics, sensitive both to properties of the nucleus and to the electronic wave functions in the nuclear region~\cite{sobelman96a}. By comparing measured and calculated values of the hyperfine structure, information about nuclear and atomic structure may be deduced. Such hyperfine comparisons play an important role in atomic parity violation (APV) studies~\cite{ginges04a, safronova18a}, by contributing to the understanding of the modeling of atomic wave functions and to the assignment of the error in the theoretical value for the APV amplitude. Atomic parity violation studies provide a sensitive and unique probe of possible new physics beyond the Standard Model, including providing a window into a possible dark sector \cite{safronova18a}.

Much of the focus on the hyperfine structure related to studies of 
atomic parity violation has been for the ground and low-lying states~\cite{dzuba02a,porsev09a}. 
In the current work, we explore the behavior of the hyperfine structure across principal quantum number $n$ to $n=18$ for heavy alkali-metal atoms. We are interested, particularly, in the contribution of the correlation corrections, that is, what remains beyond the mean-field result. We study the corrections to the states $n\,^{2}S_{1/2}$ and $n\,^{2}P_{1/2}$ -- which we refer to simply as $ns$ and $np_{1/2}$ -- for neutral alkali-metal atoms of interest for APV studies, Rb~\cite{dzuba12a}, Cs \cite{toh19a}, and Fr~\cite{gomez06a}. 

The motivation to study the hyperfine structure for high states of heavy alkali-metal atoms comes from the recent works \cite{ginges17a,ginges18a}, where it was found that: (i) the uncertainties in calculations of the hyperfine structure associated with nuclear properties are significant and hinder the extraction of information about the electronic wave functions in hyperfine comparisons \cite{ginges17a}, and; (ii) the reliance on explicit information about nuclear properties may be removed by constructing a ratio from experimentally- and theoretically-deduced hyperfine structure for states with high $n$ \cite{ginges18a}. Improving our modelling of the hyperfine structure for high states is critical for removing the crippling dependence on nuclear uncertainties and for accurately probing the electronic wave functions in the nuclear region through the ratio method \cite{ginges18a}. 

The usual starting point for accurate calculations of the hyperfine structure for heavy atoms is the relativistic Hartree-Fock approximation. 
The many-body corrections are often divided into a part arising from the distortion of the atomic core due to the external field (magnetic hyperfine interaction), referred to as {\it core polarization}, and the remaining part that we refer to as the 
{\it correlation} correction. For the alkali-metal atoms, with a single valence electron above closed shells, this correlation correction is dominated by the effect of the polarization of the atomic core by the Coulomb field of the valence electron.

There have been several studies of the trends in the core polarization and correlation corrections to the hyperfine structure in heavy alkali-metal atoms --  for the ground states across nuclear charge $Z$~\cite{vajed-samii82a, heully85a} and for $s$, $p$, and $d$ states across principal quantum number $n$ \cite{vajed-samii81a,heully82a,owusu97a,sahoo15a,gomez08a,Tang_2019}. In the latter works, the levelling of the relative core polarization corrections and the falling of the relative correlation corrections with increase in $n$ has been noted. However, only the lowest few states were considered in these works, with $\Delta n = 2-6$, and most of the studies were limited to $s$ states.

In the recent work \cite{ginges18a}, the hyperfine structure for $s$ states was evaluated for Cs, Fr, Ba$^+$, and Ra$^+$ up to $n=16$, and the relative correlation corrections were shown to fall off quickly and approach constant and non-zero values for high $n$. 
In the current work, we explore this curious behavior in more detail, and we extend the previous studies by investigating the trend in relative core polarization and correlation corrections for both $s$ and $p_{1/2}$ states for Rb, Cs, and Fr up to $n = 18$. 
We deduce relative correlation corrections from {\it measured} values for the hyperfine constants, and we demonstrate that these values agree well with our theoretical results. 

Further, the existence of the trend in relative correlation corrections allows us to make very accurate predictions of the hyperfine constants for excited states, by combining theoretical calculations for the excited states with measurements from lower ones.
We do this for the $s$ and $p_{1/2}$ states of Rb and Cs up to $n = 17$, and of Fr up to $n=12$.
For most states of Rb and Cs, we believe these results are accurate to about $0.1\%$ or better.
As a test, we also make predictions in the same way for excited states where experiments have been performed, and find excellent agreement between our predictions and the measured hyperfine constants ($0.03-0.05\%$ deviation).

This paper is organized as follows. In Section \ref{sec:hyperfine-struc} we present the basic theory for the hyperfine structure and provide further details for motivating the study of the correlation corrections for high states. In Section \ref{sec:core-pol}, the contribution of the core polarization to the hyperfine structure is evaluated, and it is shown that the relative correction is (very nearly) constant across principal quantum number for $s$ and $p_{1/2}$ states. In Section \ref{sec:corr-pot}, we describe the all-orders correlation potential method, and in Section \ref{sec:corr-trend} results for the relative correlation corrections across $n$ are obtained. The trends in the relative correlation corrections are shown to be supported by measurements of the hyperfine structure. 
Finally, in Section~\ref{sec:predictions}, we utilize the correlation trends to make high-accuracy predictions for the hyperfine constants.
Concluding remarks are presented in Section \ref{sec:conclusion}.

\section{Hyperfine structure} \label{sec:hyperfine-struc}

The interaction between the magnetic dipole moment of the nucleus and the magnetic field from unpaired electrons in the atom produces small splittings in the electronic spectra of the levels referred to as the hyperfine structure. 
The relativistic operator for the magnetic hyperfine interaction is 
\begin{equation}
\label{eq:hfs}
h_{\rm hfs}=c{\boldsymbol \alpha}\cdot {\bf A}
=\frac{1}{c}\frac{{\boldsymbol \mu}\cdot ({\boldsymbol r}\times
{\boldsymbol \alpha})}{r^3}F(r) \ ,
\end{equation}
where ${\boldsymbol \alpha}$ is a Dirac matrix, ${\bf A}$ is the nuclear 
vector potential, ${\boldsymbol \mu}=\mu {\bf I}/I$ is the nuclear magnetic 
moment, and ${\bf I}$ is the nuclear spin. $F(r)$ describes the nuclear magnetization 
distribution and $F(r) =1$ for a point-nucleus. We use atomic units throughout, $|e| = m_e = \hbar = 4\pi\epsilon_0 =1, c= 1/\alpha$, unless otherwise stated.

The magnetic hyperfine structure  (HFS) is often quantified by the hyperfine $A$ constant, which may be expressed in lowest order as   
\begin{equation}
\label{eq:hfs2}
A_{n\kappa} = -\frac{\alpha^2}{m_p}\frac{g_I\kappa}{J(J+1)} \int_0^{\infty}dr\,f(r)g(r)/r^2 \ ,
\end{equation}
where $n$ is the principal quantum number, $\kappa$ is the relativistic angular momentum quantum number, with $\kappa = -1, 1, -2, ...$ for $s$, $p_{1/2}$, $p_{3/2}$, ..., etc., $J$ is the 
electronic angular momentum, $m_p$ is the proton mass, and $g_I = \mu/(\mu_N I)$ is the nuclear g-factor.
Here, $f$ and $g$ are the upper and lower radial components of the single-particle Dirac orbitals
\[
\varphi_{n\kappa m}(\boldsymbol{r}) = \frac{1}{r}
\begin{pmatrix}f_{n\kappa}(r) \, \Omega_{\kappa m}(\boldsymbol{\hat r}) \\ 
i \alpha g_{n\kappa}(r)\,  \Omega_{-\kappa m}(\boldsymbol{\hat r})\end{pmatrix},
\]
($\Omega$ is a spherical spinor) which satisfy the relativistic Hartree-Fock (RHF) equations
\begin{equation}
\label{eq:dirac}
\big(c{\boldsymbol \alpha}\cdot {\bf p}+(\beta-1)c^2 + V_{\rm
  nuc}(r)+V_{\rm HF}\big) \varphi = \epsilon \varphi \ ,
\end{equation}
where $\beta$ is a Dirac matrix, and $V_{\rm  HF}$ and $V_{\rm nuc}$ are the Hartree-Fock and nuclear potentials, respectively. 
See, e.g., Ref.~\cite{johnson07a} for detailed expressions for $V_{\rm HF}$.
We use the Fermi distribution to form the nuclear potential, with the thickness parameter corresponding to the 90\%--10\% density fall-off set to 2.3\,fm, 
and the half-density radius found from the root-mean-square charge radii tabulated in Ref.~\cite{angeli13a}.

An accurate theoretical description of the hyperfine structure goes beyond the expression presented in Eq.~(\ref{eq:hfs2}). 
The largest corrections are due to many-body effects, and to reach an accuracy within $\sim 1\%$ or $\sim 0.1\%$,  finite-nucleus magnetization and quantum electrodynamic (QED) 
radiative corrections must be included.
In the recent paper~\cite{ginges18a}, the following parametrization for the hyperfine constant was introduced, 
\begin{equation}
\label{eq:ABWQED}
A_{n\kappa}=A_{n\kappa}^{\rm MB}\frac{\mu}{\mu_N}\Big(1+\frac{\alpha}{\pi} F^{\rm
  BW}_{n\kappa}+\frac{\alpha}{\pi} F^{\rm QED}_{n\kappa}\Big) \ ,
\end{equation}
which conveniently factors out the nuclear and QED radiative corrections. 
We will adopt the same parametrization in this work. 
The first term on the right-hand-side $A_{n\kappa}^{\rm MB}$ corresponds to an electronic many-body value found with
$\mu = \mu_N$, point-nucleus magnetization ($F(r) = 1$), and no QED corrections.
$F^{\rm BW}_{n\kappa}$ is the relative Bohr-Weisskopf correction originating 
from the finite magnetization distribution of the nucleus and  $ F^{\rm QED}_{n\kappa}$
is the relative QED radiative correction. Note that $A_{n\kappa}^{\rm MB}$ contains the 
nuclear spin $I$, from the nuclear g-factor $g_I$ --  see Eq.~(\ref{eq:hfs2}) -- and so 
may be different for different isotopes.

The QED radiative corrections to the hyperfine structure for low-lying states of heavy atoms were rigorously evaluated 
in ~\cite{sapirstein03a,sapirstein06a,sapirstein08a,ginges17a}, and for $s$ states 
across principal quantum number in~\cite{ginges18a}. 
A detailed study of the Bohr-Weisskopf effect in different nuclear models was carried out recently for heavy atoms~\cite{ginges17a}, 
and across principal quantum number in Ref.~\cite{ginges18a}. In the current work, where Bohr-Weisskopf contributions are required, we use the results 
of Refs.~\cite{ginges17a,ginges18a} found in the nuclear single-particle model (see, e.g., Ref.~\cite{volotka08a}).

In Ref.~\cite{ginges18a}, it was demonstrated theoretically that the relative Bohr-Weisskopf and QED radiative corrections, $F_{n\kappa}^{\rm BW}$ and $F_{n\kappa}^{\rm QED}$, are independent of principal quantum 
number $n$ to high accuracy for $s$ and $p$ states for heavy atoms of interest for APV studies \cite{ginges18a}. And indeed, the $n$-independence of the Bohr-Weisskopf effect has been experimentally 
demonstrated across $n=5-7$ for Rb \cite{galvan07a}. What this means is that the nuclear and QED terms that appear as a factor in Eq.~(\ref{eq:ABWQED}) are the same for all $n$, and that they may be determined from 
a ratio of measured and calculated values for the hyperfine structure, $A^{\rm exp}_{n'\kappa}/A^{\rm MB}_{n'\kappa}$ \cite{ginges18a}. This may be readily seen from Eq.~(\ref{eq:ABWQED}) by considering 
the equation for two principal quantum numbers $n$ and $n'$, and setting the total hyperfine constant for the state $n'\kappa$ to the measured value $A_{n'\kappa}^{\rm exp}$ when all effects are included correctly. 

Remarkably, removing the explicit dependence on nuclear structure  makes it possible to probe the electronic wave functions in the nuclear region with greatly-improved sensitivity (potentially testing the many-body theory at the level of 0.1\% or better) compared to what could be 
possible from a direct hyperfine comparison. For example, the nuclear magnetic moment alone for isotopes of Fr has an uncertainty $\sim 1\%$, limiting this comparison.
Note that the preference to determine the ratio described above for the {\it highly excited states} comes from the observation that the relative correlation corrections are significantly smaller for the higher states compared to the ground or low-lying states, as 
we will see in more detail later. Because it is expected that the uncertainty in the many-body calculation of $A^{\rm MB}_{n\kappa}$ is related to the size of the correlation correction, then it is anticipated that 
higher accuracy may be achieved for the high-lying states compared to the lower ones. 
This motivates our study of the hyperfine constant across principal quantum number, as we strive to better understand and 
evaluate it. 

The subject of the current work relates to the many-body term $A_{n\kappa}^{\rm MB}$,
and in particular, the trends arising from contributions beyond the RHF approximation.
The electronic term may be approximated by 
\begin{equation}
A^{\rm MB}_{n\kappa} \approx  A^{\rm HF}_{n\kappa}(1+F^{\delta V}_{n\kappa})(1+F^{\Sigma}_{n\kappa}),
\end{equation}
where $A^{\rm HF}_{n\kappa}$ is the result at the relativistic Hartree-Fock level of 
approximation, $F^{\delta V}_{n\kappa}$  is a relative correction arising due to 
polarization of the atomic core by the hyperfine field, and $F^{\Sigma}_{n\kappa}$ is the relative correction 
arising from valence-core electron correlations.
In the following sections we study the relative core polarization and correlation corrections 
across principal quantum number $n$. We will see that the core polarization correction is (nearly) 
independent of principal quantum number, and that the relative correlation corrections drop 
quickly and approach a constant value for high $n$.

\section{Core polarization} \label{sec:core-pol}

One of the dominant many-body corrections to the hyperfine structure arises due to polarization 
of the atomic core by the hyperfine field -- core-polarization.
We include this in our calculations using the time-dependent Hartree-Fock method, which is equivalent to the random 
phase approximation (RPA) with exchange. Effectively, it modifies the hyperfine operator, such that~\cite{dzuba87a}
\begin{equation}
 \label{eq:rpa_operator}
h_{\rm hfs} \rightarrow h_{\rm hfs} + \delta V_{\rm hfs}~.
\end{equation}
Only the exchange term contributes to the core polarization for the magnetic hyperfine structure, and so it is 
sometimes referred to as the ``exchange core polarization''.

To calculate $\delta V_{\rm hfs}$, we use the time-dependent Hartree-Fock (TDHF) method, in which the single-particle orbitals are expressed
\begin{equation}
\varphi=\varphi^{(0)} + \delta\varphi,
\end{equation}
where $\varphi^{(0)}$ is the unperturbed orbital, and $\delta\varphi$ is the correction due hyperfine interaction.
Then, the set of TDHF equations
\begin{align}
(h^{\rm HF} - \varepsilon_c) \delta\varphi_c &=-(h_{\rm hfs} + \delta V_{\rm hfs}-\delta\varepsilon_c)\varphi^{(0)}_c
\label{eq:tdhf}
\end{align}
is solved self-consistently for all the core orbitals.
Here, the index $c$ denotes a state in the core, $\delta\varepsilon=\bra{\varphi^{(0)}_c}{h_{\rm hfs} + \delta V_{\rm hfs}}\ket{\varphi^{(0)}_c}$ is the correction to the energy for the core orbital $c$, and $h^{\rm HF}$ is the single-particle Hamiltonian operator on the left-hand side of Eq.~(\ref{eq:dirac}).

\begin{figure}
\includegraphics[width=\columnwidth]{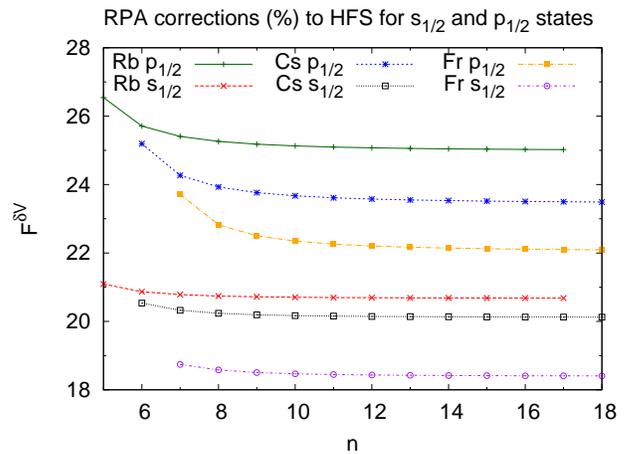}
\caption{The relative core polarization (RPA) correction $F^{\delta V}$ in \% for $s_{1/2}$ and $p_{1/2}$ states of Rb, Cs, and Fr.
  \label{fig:RPA-all-corr}}
\end{figure}

In Fig.~\ref{fig:RPA-all-corr} we plot the relative core polarization corrections
($F^{\delta V}$) as a function of $n$ for $s_{1/2}$ and $p_{1/2}$ states
 of Rb, Cs, and Fr. The correction $F^{\delta V}$ is defined as 
$\langle \varphi |\delta V_{\rm hfs}| \varphi \rangle /\langle \varphi |h_{\rm hfs}| \varphi \rangle $, found from the ratio of the 
hyperfine constant with core polarization included to that without.  
We observe the following trends: (i) the relative 
  core polarization corrections are constant for higher states; (ii) the corrections decrease 
  with increasing $Z$. These observations are in agreement with previous studies. For example, the first point has been noted in 
Refs.~\cite{heully82a,vajed-samii81a,owusu97a} where the first few $s$ states of K, Rb, and Fr were considered, and the second 
point has been noted in Refs.~\cite{heully85a, vajed-samii82a}.

In the current work we extend previous studies by going to higher $n$, and 
by studying the trend for the $p_{1/2}$ states. 
The relative core polarization corrections decrease slightly with increase in $n$ for the lowest-lying levels, and they approach constant values as $n$ is further increased. 
For $s_{1/2}$ states, they level out at 21\%, 20\%,  and 18\% for Rb, Cs, and Fr, respectively.  
The relative core polarization corrections for $p_{1/2}$ states are significantly larger than for $s_{1/2}$ states, and they approach 25\%, 24\%, and 22\% for Rb, Cs, and Fr. 

\section{Correlation potential method} \label{sec:corr-pot}

\begin{figure}[bth]
\includegraphics[width=0.8\columnwidth]{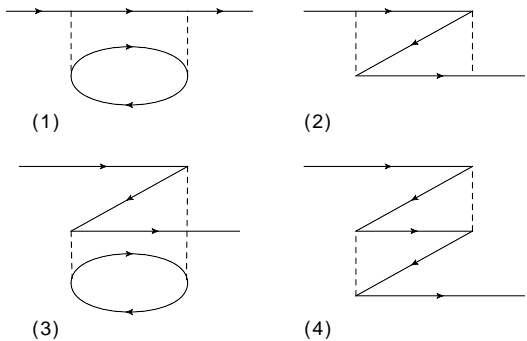}
\caption{Goldstone diagrams for the second-order  
correlation corrections to the energy, where (1) and (3) are direct terms, and (2) and (4) are exchange terms. Solid forward lines correspond to electron lines, backward lines to holes, dashed lines to Coulomb lines.
\label{fig:second}}
\end{figure}

We use the correlation potential method~\cite{dzuba87a} to include valence-core electron correlations. In this method, a correlation potential $\Sigma({\bf r}_i, {\bf r}_j, \epsilon)$ is added to the RHF equations (Eq.~(\ref{eq:dirac})), and new orbitals $\varphi_{\rm Br}$ (Brueckner orbitals) and Brueckner energies are obtained for the valence states.
The correlation potential is defined such that the average value of the second-order  correlation potential corresponds to the second-order correlation correction to the energy.  Diagrams are presented in Fig.~\ref{fig:second}.

We use the Feynman diagram technique to include higher-order correlations, and a fitting procedure is used to 
approximate the inclusion of missed diagrams.
Calculation of  the hyperfine constant with valence-core correlations included, along with core polarization, corresponds to evaluation of the matrix element  
\begin{equation}
 \langle \varphi_{\rm Br}|h_{\rm
  hfs}+\delta V_{\rm hfs}| \varphi_{\rm Br}\rangle~.
\end{equation}
In the following sections we describe how higher orders are included in the potential.

\subsection{Higher orders}

\begin{figure}[bth]
\includegraphics[width=1.0\columnwidth]{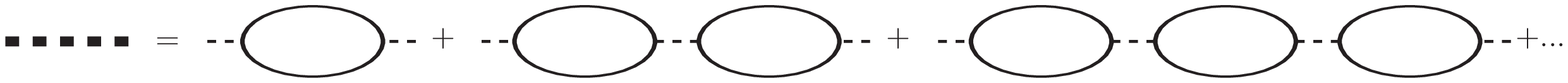}
\caption{All-orders electron-electron screening of the Coulomb interaction, corresponding to a series of hole-particle loops, producing a dressed Coulomb line.
    \label{fig:screened-coulomb}}
\end{figure}

\begin{figure}[bth]
\includegraphics[width=1.0\columnwidth]{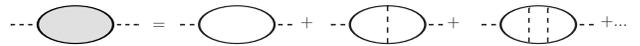}
\caption{All-orders summation of hole-particle interaction, forming a dressed polarization loop.  
    \label{fig:hole-particle}}
\end{figure}

\begin{figure}[bth]
\includegraphics[width=0.4\columnwidth]{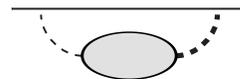}
\caption{Feynman diagram for the direct, all-orders correction to the energy. Note that in the lowest order, this diagram is equivalent to the sum of Goldstone diagrams (1) and (3) of Fig.~\ref{fig:second}.
    \label{fig:all-order-dir}}
\end{figure}

The Feynman diagram technique is used to re-express the Goldstone diagrams in Fig.~\ref{fig:second}. 
Then it is relatively straight-forward to include certain classes of diagrams to all-orders in the Coulomb interaction. We do this using the method developed in Ref.~\cite{dzuba88a}; we refer the reader to that work for the relevant equations.
The most important class of diagrams corresponds to electron-electron screening of the Coulomb interaction, represented in Fig.~\ref{fig:screened-coulomb}. 
Another class of diagrams -- hole-particle interaction -- is included through dressed hole-particle loops, as depicted in Fig.~\ref{fig:hole-particle}. All-orders electron-electron screening and the hole-particle interaction are included in the Feynman diagram for the direct terms, as shown in Fig.~\ref{fig:all-order-dir}. 

Exchange diagrams are usually considered to be small in comparison to direct diagrams (see, e.g., Ref.~\cite{dzuba08a}), and a simpler approach is used. The exchange part of the correlation potential is evaluated in the second order. This calculation involves a sum over a complete set of states in the internal lines, and to discretize the states in this sum, we introduce a cavity of radius $40\,{a}_{B}$ and diagonalize the relativistic Hartree-Fock Hamiltonian on a set of 40 splines of order $k=9$ \cite{johnson07a}. Higher order correlations are accounted for by using multipolarity-dependent screening factors $f_k$, where $k$ is the multipolarity of the Coulomb interaction. These rescale the Coulomb integrals $g_k$ that correspond to the Coulomb lines in the exchange diagrams (2) and (4) of Fig.~\ref{fig:second} as $f_k g_k$. The screening factors $f_k$ are found from the ratio \cite{dzuba08a}
\begin{equation}
f_k = \langle v| {\Sigma}^{\infty,ee}_{{\rm dir},k}|v \rangle/\langle v|{\Sigma}^{(2)}_{{\rm dir},k}|v \rangle,
\end{equation}
where ${\Sigma}^{\infty,ee}_{{\rm dir},k} $ and ${\Sigma}^{(2)}_{{\rm dir},k} $ refer to the direct parts of the all-order and second-order correlation potentials, respectively. These factors include only the dominant electron-electron screening correction.

To improve the accuracy of our calculations further, we multiply the correlation potential by a fitting factor which is tuned to reproduce the experimental binding energies. Beyond the higher-order corrections that may be absorbed into a correlation potential, there are other small corrections (at the level of 1\% or less). This includes so-called structural radiation and normalization of states~\cite{dzuba87a}. These are taken into account in this work. 

We also include the Breit correction, accounting for retardation and magnetic effects. 
The effective Breit Hamiltonian, 
\begin{equation}
h^{\rm B}(\boldsymbol{r}_1,\boldsymbol{r}_2)=\frac{-1}{2r_{12}}\left(\boldsymbol{\alpha}_1\cdot\boldsymbol{\alpha}_2+\frac{(\boldsymbol{\alpha}_1\cdot\boldsymbol{r}_{12})(\boldsymbol{\alpha}_2\cdot\boldsymbol{r}_{12})}{r_{12}^2}\right),
\end{equation}
where $\boldsymbol{r}_{12}=\boldsymbol{r}_1-\boldsymbol{r}_2$,
is included in the Hartree-Fock equations self-consistently at the RHF and RPA levels.

\section{Correlation corrections across principal quantum number} \label{sec:corr-trend}

In our previous work~\cite{ginges18a} we observed that the relative correlation corrections for $s$ states of Cs, Fr, Ba$^+$, and Ra$^+$ tend towards constant and non-zero values for high $n$. 
We define the relative correlation correction $F^{\Sigma}$ as
\begin{equation}
F^{\Sigma}=(A 
- A^{\rm RPA}) /A^{\rm RPA} \,, \label{eq:sigma-corr}
\end{equation}
where $A$ is the total hyperfine constant and $A^{\rm RPA}$ is the result of our calculation at the RPA level, including Bohr-Weisskopf and QED corrections using Eq.~(\ref{eq:ABWQED}). 
The relative correlation correction describes the correlations included beyond the mean field approximation. Note that in the {\it theoretical} evaluation of $F^{\Sigma}$ any dependence on nuclear 
properties and QED radiative corrections in $A$ and $A^{\rm RPA}$ factors out, as long as these corrections are treated in the same way. It's also possible to find {\it experimental}  values for the 
relative correlation corrections, as done below, using measured values for the hyperfine constants $A$ in Eq.~(\ref{eq:sigma-corr}).

\subsection{Theory}

 \begin{figure}[tb]
\includegraphics[width=\columnwidth]{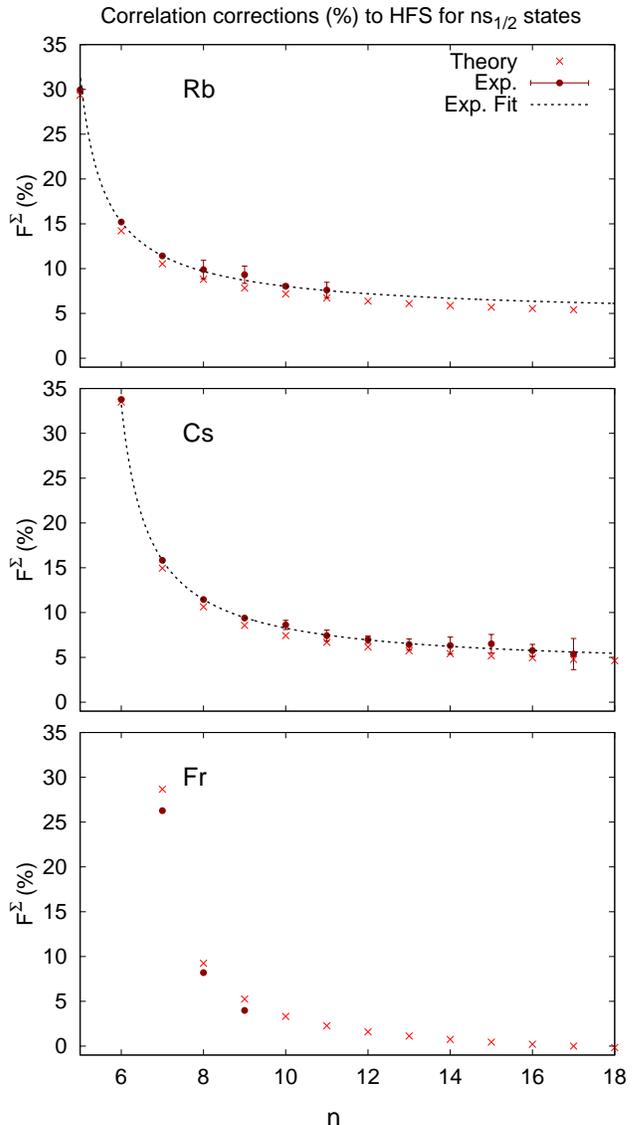}
\vspace{-0.8cm}
\caption{Relative correlation corrections $F^{\Sigma}$ for $ns$ states of $^{87}$Rb, $^{133}$Cs, and $^{211}$Fr. 
All-order many-body results are shown as crosses, values extracted from measurements (see Tables in the following sections) are shown as circles with error bars, and the dashed lines 
correspond to least squares fits to the measured data. 
Uncertainties from the choice of nuclear magnetic moments and Bohr-Weisskopf corrections are not included; these may be sizeable for Fr.
Measured values of hyperfine constants for $^{210}$Fr were simply rescaled using $g_I^{211}/g_I^{210}$.
\label{fig:rel-corr-s}}
\end{figure}

 \begin{figure}[tbh]
\includegraphics[width=\columnwidth]{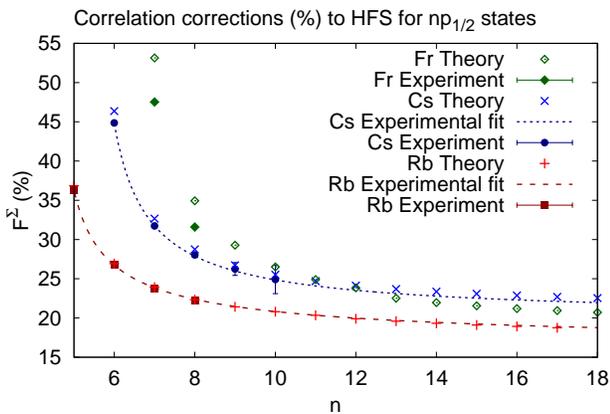}
\vspace{-1cm}
\caption{Relative correlation corrections $F^{\Sigma}$ for $np_{1/2}$ states for $^{87}$Rb, $^{133}$Cs, and $^{211}$Fr. See caption for Fig.~\ref{fig:rel-corr-s} for further explanation. Note that for Fr there is no available data 
for the QED corrections, and so these are not included in extracting the measured values for this atom.
The measured value for Fr 8$p_{1/2}$ was found by simply rescaling the hyperfine constant of $^{212}$Fr by $g_I^{211}/g_I^{212}$.
}
\label{fig:rel-corr-p}
\end{figure}

The results of our many-body calculations for the relative correlation corrections to the hyperfine structure for Rb, Cs, and Fr are presented in Fig.~\ref{fig:rel-corr-s} for $s$ states across principal quantum $n$ from the ground state to $n = 18$.
The theory results are shown as crosses in the figure (the values are presented in tables in the following sections).
In agreement with Refs.~\cite{vajed-samii82a,heully85a}, our results for the relative correlation corrections to the hyperfine structure for the ground state of Cs have a larger value than for the ground states of Rb and Fr. The relative correlation corrections exhibit a strong decrease at low $n$ and level off to constant, non-zero values for high $n$.  The decrease in the correlation corrections with principal quantum number $n$ has been observed theoretically in several alkali-metal atoms, see for example Refs.~\cite{vajed-samii81a,owusu97a,sahoo15a,gomez08a,ginges18a}. The references~\cite{vajed-samii82a,vajed-samii81a} provide an explanation for these trends  -- that they arise due to competing factors, the electric polarizability of the atom and the separation distance between the core and valence electron. 
Indeed, at large distances, the correlation potential approaches a local polarization potential that involves the electric polarizability of the atomic core $\alpha$, $\Sigma_{r\rightarrow \infty} \approx -\alpha /(2r^4)$. 
 The decrease in the relative correlation potential with increase in $n$ has been seen previously for the lowest states, and observed over a higher range in $n$, up to $n=16$, more recently in Ref.~\cite{ginges18a}. In the current work, we study the trend up to $n =18$, and present the first results for Rb across a wide range in $n$. We demonstrate that for the three atoms, $F^{\Sigma}$ tends towards constant and non-zero values with increase in $n$. At $n=18$, we find that the relative correlation corrections are 5.2\%, 4.5\% , and $-0.3$\% for Rb, Cs, and Fr. 
Note that for Fr, this value is negative, an interesting result that has been observed previously for Ra$^+$ \cite{owusu97a}, which is consistent with our Ra$^+$ results \cite{ginges18a}.

We have also calculated the relative correlation corrections to $np_{1/2}$ states up to $n = 18$ for Rb, Cs, and Fr. Our results are shown in Fig.~\ref{fig:rel-corr-p}. It is seen that the relative correlation corrections are significantly larger for the $p_{1/2}$ states compared to the $s_{1/2}$ states. This is consistent with the results of Ref.~\cite{heully85a}, where calculations were performed for the lowest-lying $p_{1/2}$ states. The $p_{1/2}$ states exhibit similar behaviour to the $s$ states for $F^{\Sigma}$, with the values decreasing quickly at low $n$ as $n$ is increased, and leveling off to constant and non-zero values for high $n$. 

Simple analytical arguments support the observation that the relative correlation corrections tend towards constant values. These are applicable for both $s$ and $p_{1/2}$ states. The Brueckner orbitals may be written as $\varphi_{\rm Br} = \varphi + \delta \varphi $, where $ \delta \varphi = \sum _{m} |\varphi_m\rangle \langle \varphi _m | \Sigma^{(\infty)} | \varphi \rangle/(\epsilon -\epsilon_m) $ and $\varphi$ and $\epsilon$ correspond to RHF values; see Eq.~(\ref{eq:dirac}). 
The correlation correction to the hyperfine constant $A$ may then be expressed as $\Delta A \propto \langle \varphi | h_{\rm hfs} | \delta \varphi \rangle +  \langle \delta \varphi | h_{\rm hfs} | \varphi \rangle$.
We will proceed by considering the simple case where the sum is dominated by a single term where $\epsilon \approx \epsilon_m$.
It is known that $\langle \varphi| h_{\rm hfs} | \varphi\rangle \propto 1/\nu^3$~\cite{sobelman96a}, where $\nu$ is the effective principal quantum number, and the energy interval may be approximated as $1/\nu^2 - 1/\nu_m^2 \propto 1/\nu^3$
for large $\nu$ (since $\nu \approx \nu_m$ and the difference between the two is approximately a constant). 
Finally, $\langle \varphi|\Sigma^{(\infty)}| \varphi \rangle$ is simply the correlation correction to the energy, which also scales as $1/\nu^3$. Using these scalings, it is then seen that 
$\Delta A \propto 1/\nu^3 $.
Therefore, the \textit{relative} correlation correction tends towards a constant for high $n$.

\subsection{Experiment} \label{sec:corr-exp}

In this section we look at the trends in the relative correlation corrections derived from experiment and we use the combination of our calculations with these experimental results to make high-accuracy predictions for the hyperfine constants.

For $^{133}$Cs, hyperfine constants have been measured for $s$ states up to $n = 17$, and there are data for several $p_{1/2}$ states. There are also a number of measurements of the hyperfine constants for $^{87}$Rb, for $s$ states up to $n=11$ and for several $p_{1/2}$ states.
 For Fr, the data is more limited. Measurements include the hyperfine constants for $7s$ and $7p_{1/2}$ states of $^{211}$Fr \cite{grossman99a},  the $8s$ and $9s$ states of the isotope $^{210}$Fr, and the $8p_{1/2}$ state of $^{212}$Fr. 
The relevant measured data for $^{133}$Cs, $^{87}$Rb and $^{210,212}$Fr are presented in the following sections.

In Figures \ref{fig:rel-corr-s} and \ref{fig:rel-corr-p} we present these measured hyperfine constants as relative correlation corrections $F^{\Sigma}$ alongside our theory values. Indeed, it is possible to derive these relative corrections from the measured data using Eq.~(\ref{eq:sigma-corr}). These corrections are defined by what remains beyond our mean-field results, in this case our RPA values. They are found by subtracting $A^{\rm RPA}$, which include the nuclear magnetic moment and the Bohr-Weisskopf and QED radiative corrections. For reference, and illustration of the procedure, we present the results of our calculations at the RHF and RPA levels (without Bohr-Weisskopf or QED corrections) for the states up to $n = 12$ in Table~\ref{tab:RHF-RPA}. This data, along with the values for $F^{\rm BW}$ and $F^{\rm QED}$ presented in Table~\ref{tab:BW-QED}, are used to determine $A^{\rm RPA}$ according to Eq.~(\ref{eq:ABWQED}).

\begin{table}[tb]
\caption{$A^{\rm MB}\mu/\mu_N$ for $s$ and $p_{1/2}$ states of $^{87}$Rb, $^{133}$Cs, and $^{211}$Fr as calculated at the RHF and RPA levels. Bohr-Weisskopf and QED radiative corrections are not included.
Units: MHz. 
\label{tab:RHF-RPA}}
 \begin{ruledtabular}
  \begin{tabular}{cccccccccc}
  &\multicolumn{2}{c}{$^{87}$Rb}&\multicolumn{2}{c}{$^{133}$Cs}&\multicolumn{2}{c}{$^{211}$Fr} \\ \cline{2-3} \cline{4-5} \cline{6-7}
 & RHF & RPA & RHF & RPA & RHF & RPA \\ \hline
 5$s$	&	2183	&	2644	&		&		&		&		\\
6$s$	&	583.1	&	704.7	&	1434	&	1728	&		&		\\
7$s$	&	238.8	&	288.5	&	393.9	&	474.0	&	5929	&	7040	\\
8$s$	&	120.6	&	145.6	&	164.5	&	197.8	&	1520	&	1802	\\
9$s$	&	69.24	&	83.59	&	84.11	&	101.1	&	624.0	&	739.5	\\
10$s$	&	43.37	&	52.35	&	48.71	&	58.53	&	316.8	&	375.3	\\
11$s$	&	28.95	&	34.94	&	30.70	&	36.89	&	182.7	&	216.5	\\
12$s$	&	20.27	&	24.47	&	20.59	&	24.74	&	114.9	&	136.1	\\
5$p_{1/2}$	&	236.8	&	299.6	&		&		&		&		\\
6$p_{1/2}$	&	83.21	&	104.6	&	161.0	&	201.6	&		&		\\
7$p_{1/2}$	&	38.60	&	48.41	&	57.65	&	71.64	&	628.2	&	777.2	\\
8$p_{1/2}$	&	20.97	&	26.27	&	27.09	&	33.57	&	222.9	&	273.8	\\
9$p_{1/2}$	&	12.64	&	15.82	&	14.85	&	18.38	&	104.4	&	127.9	\\
10$p_{1/2}$	&	8.192	&	10.25	&	9.002	&	11.13	&	57.13	&	69.90	\\
11$p_{1/2}$	&	5.610	&	7.018	&	5.863	&	7.248	&	34.61	&	42.31	\\
12$p_{1/2}$	&	4.008	&	5.013	&	4.030	&	4.980	&	22.53	&	27.53	\\
\end{tabular}
\end{ruledtabular}
\end{table}

 \begin{table}[b]
  \caption{Relative Bohr-Weisskopf corrections, $F^{\rm BW}$, in the nuclear single-particle model at the RPA level 
 and relative QED radiative corrections, $F^{\rm QED}$, in the core-Hartree approximation 
 for the lowest $s$ and $p_{1/2}$ states of $^{87}$Rb, $^{133}$Cs, and $^{211}$Fr. The values for $F^{\rm BW}$ for $p_{1/2}$ states of $^{87}$Rb and $^{211}$Fr are results of the current work.
  }
 \label{tab:BW-QED}
 \begin{ruledtabular}
  \begin{tabular}{cccccc}
 	     &$\mu$~\cite{Stone2005} & \multicolumn{2}{c}{$F^{\rm BW}$} & \multicolumn{2}{c}{$F^{\rm QED}$} \\ \cline{3-4} \cline{5-6}
 	     &		&	$s$	&	$p_{1/2}$ & $s$~\cite{ginges17a} & $p_{1/2}$~\cite{sapirstein06a} \\ \hline
 $^{87}$Rb& 2.751818(2) 		& 	-1.20\tm[1]  &  	0.0098		&	-1.039 &	-0.023	\\
 $^{133}$Cs&  2.582025(3)		& 	-0.898\tm[2] &  	-0.056\tm[2]		&	-1.638 & -0.096	\\
  $^{211}$Fr&  	4.00(8)	        & 	-5.6\tm[1] &  	-1.69		&	-2.59	& -		\\
  \end{tabular}
 \end{ruledtabular}
 \tablenotetext[1]{Reference~\cite{ginges17a}}
 \tablenotetext[2]{Reference~\cite{ginges18a}}
 \end{table}

The Bohr-Weisskopf corrections, $F^{\rm BW}$, and QED radiative corrections, $F^{\rm QED}$, are presented in Table~\ref{tab:BW-QED}. The Bohr-Weisskopf results were obtained in the single-particle model at the RPA level~\cite{ginges17a,ginges18a}. The results for $^{87}$Rb and $^{211}$Fr $p_{1/2}$ states were found in the current work.  For the QED radiative corrections, $F^{\rm QED}$, we use the results of Refs~\cite{ginges17a,sapirstein06a}. The QED correction for the $p_{1/2}$ state of Fr has not been determined.

The relative correlation corrections may be extracted from measured data for the hyperfine constants as follows. The measured hyperfine constant for Cs $9s$ is $A^{\rm exp} = 109.93(9)$\,MHz~\cite{stalnaker10a}. Our theory result at the RPA level is $A^{\rm MB}\mu/\mu_N = 101.09$ MHz, and from the data in Table~\ref{tab:BW-QED} for $F^{\rm BW}$ and $F^{\rm QED}$, we obtain $A^{\rm RPA}=100.50$\,MHz using Eq.~(\ref{eq:ABWQED}). The relative correlation correction for Cs $9s$ is then found from Eq.~(\ref{eq:sigma-corr}), giving $F^{\Sigma} = 9.39(9)\%$.   

It is seen from Figs.~\ref{fig:rel-corr-s} and \ref{fig:rel-corr-p}  that our calculated $F^\Sigma$ has generally excellent agreement with experiment for the considered $s$ and $p_{1/2}$ states. 
The agreement is particularly good for the ground states of Rb and Cs and the $p_{1/2}$ states of Rb.
Note that the error bars for $F^{\Sigma}$ only include the uncertainties associated with the measured hyperfine constants, and they do not include uncertainties associated with extracting the relative correlation corrections, such as from the nuclear magnetic moment. The nuclear magnetic moment of Fr in particular has a very large uncertainty ($2\%$), which may explain the difference in the results between theory and experiment.

The smooth dependence of the relative correlation corrections on principal quantum number is supported by the measured values.
This allows us to make highly accurate predictions of the hyperfine constants, as implemented in the following section.

\section{Predictions}\label{sec:predictions}

While there is a known trend of $1/\nu^3 $ for the hyperfine constants, more accurate predictions may be found by taking advantage of the observed trend in the relative correlation corrections.
We do this using two methods: by fitting the available experimental data to the theoretically-motivated trend in the relative correlation corrections, and by using the ratio method, as developed in Ref.~\cite{ginges18a}.
We note that the observed trends, from theory and experiment, align most strongly for the excited states, and so the ground state is to be treated separately (accurate calculations of the ground state hyperfine splitting for these atoms have been presented recently in Ref.~\cite{ginges17a}).

\subsection{Fit method}

We fit the function $a(n-b)^{-c}+d$ to the relative correlation corrections ($F^\Sigma$) using weighted least squares, where $n$ is the principal quantum number and $a,b,c$ and $d$ are the fitting parameters. 
First, the parameters $b$ and $c$ are fixed by fitting the function to the theoretical $F^\Sigma$ values [$(n-b)>1$ may be thought of as an effective principal quantum number, and $c\simeq1$ determines the degree of the relative correlation trend].
Then, the $a$ and $d$ parameters are fitted to the $F^\Sigma$ values derived from experimental $A$ values; fitting these terms to the experimental data accounts for small errors in the calculated correlation corrections.
This method allows us to make accurate predictions using the fit when only few experimental values are known.
We stress that we always exclude the experimental value for the state we are making the prediction for from the fit.

Using this method we predict the hyperfine constants up to $n=17$.
These are presented for Rb and Cs in Tables~\ref{tab:fittedRb} and \ref{tab:fittedCs}, respectively, wherever the experimental uncertainty drops below 0.1\% and where measured values are currently unavailable. 
We estimate the uncertainties in the predicted values from the uncertainties in the fit parameters; the fit is mostly sensitive to the $b$ and $c$ parameters (determined from the fit to theory), and it is these which typically dominate the uncertainty.
Our predictions from the fit method all lie well within the experimental error bars.

 \begin{table*}[tb]
   \caption{Rb hyperfine constants $A$ (in MHz) given by theory and experiment, as well as the predictions from the fit and ratio methods.}
  \label{tab:fittedRb}
 \begin{ruledtabular}
  \begin{tabular}{ccccccccc}
  &  \multicolumn{4}{c}{$ns$} &  \multicolumn{4}{c}{$np_{1/2}$} \\  
$n$  &  Theory & Exp~\cite{arimondo77a} &  Fit  &  Ratio  &  Theory & Exp~\cite{arimondo77a}  &  Fit  &  Ratio  \\ \cline{2-5}\cline{6-9} 
5 	&  3401.8 	&  3417.341\ldots   	&  -   	&  -   	&  410.06 	&  406.2(8)   	&  -   	&  - \\
6 	&  800.78 	&  807.66(8)\tm[1]   	&  -   	& 807.27(61)	&  132.77 	&  132.56(3)   	&  -   	& 132.60(14)\\
7 	&  317.19 	&  319.759(28)\tm[2]   	&  -   	& 319.92(24)	&  59.996 	&  59.92(9)   	&  59.96(11)   	& 59.90(4)\\
8 	&  157.62 	&  159.2(15)   	&  158.92(65)   	& 158.94(17)	&  32.138 	&  32.12(11)   	&  32.15(9)   	& 32.09(4)\\
9 	&  89.662 	&  90.9(8)   	&  90.38(36)   	& 90.41(10)	&  19.207 	&  -   	&  19.21(6)   	& 19.18(3)\\
10 	&  55.821 	&  56.27(12)   	&  56.26(22)   	& 56.29(4)	&  12.383 	&  -   	&  12.39(5)   	& 12.36(2)\\
11 	&  37.092 	&  37.4(3)   	&  37.38(14)   	& 37.40(3)	&  8.4453 	&  -   	&  8.45(4)   	& 8.432(17)\\
12 	&  25.891 	&  -   	&  26.23(17)   	& 26.11(2)	&  6.0101 	&  -   	&  6.01(3)   	& 6.001(18)\\
13 	&  18.782 	&  -   	&  19.03(12)   	& 18.94(2)	&  4.4295 	&  -   	&  4.43(2)   	& 4.422(16)\\
14 	&  14.055 	&  -   	&  14.24(9)   	& 14.17(1)	&  3.3577 	&  -   	&  3.36(2)   	& 3.352(14)\\
15 	&  10.791 	&  -   	&  10.93(7)   	& 10.88(1)	&  2.6055 	&  -   	&  2.61(1)   	& 2.601(12)\\
16 	&  8.4646 	&  -   	&  8.57(6)   	& 8.535(6)	&  2.0623 	&  -   	&  2.06(1)   	& 2.059(10)\\
17 	&  6.7616 	&  -   	&  6.85(4)   	& 6.818(5)	&  1.6599 	&  -   	&  1.66(1)   	& 1.657(9)\\
  \end{tabular}
 \end{ruledtabular}
 \tablenotetext[1]{Reference~\cite{galvan07a}}
  \tablenotetext[2]{Reference~\cite{chui05a}}
 \end{table*}

 \begin{table*}[tb]
\caption{Cs hyperfine constants $A$ (in MHz) given by theory and experiment, as well as the predictions from the fit and ratio methods.    }
\label{tab:fittedCs}
\begin{ruledtabular}
\begin{tabular}{ccccccccc}
&  \multicolumn{4}{c}{$ns$} &  \multicolumn{4}{c}{$np_{1/2}$} \\  
$n$  &  Theory & Exp~\cite{arimondo77a} &  Fit  &  Ratio  &  Theory & Exp~\cite{arimondo77a}  &  Fit  &  Ratio \\  \cline{2-5} \cline{6-9}
6 	& 2293.3 	& 2298.157\ldots 	& - 	& - 	& 294.96 	& 291.9309(12)\tm[5] 	& - 	& -\\
7 	& 541.65 	& 545.818(16)\tm[1] 	& - 	& 545.67(40)	& 95.015 	& 94.40(5)\tm[6] 	& - 	& 94.49(26)\\
8 	& 217.51 	& 219.125(4)\tm[2] 	& - 	& 219.18(16)	& 43.209 	& 42.97(10) 	& 42.95(9) 	& 42.93(7)\\
9 	& 109.10 	& 109.93(9)\tm[3] 	& 109.98(44) 	& 109.92(8)	& 23.276 	& 23.19(15) 	& 23.16(7) 	& 23.13(4)\\
10 	& 62.505 	& 63.2(3) 	& 62.99(24) 	& 62.98(6)	& 13.968 	& 13.9(2) 	& 13.91(5) 	& 13.88(4)\\
11 	& 39.126 	& 39.4(2) 	& 39.42(15) 	& 39.42(3)	& 9.0343 	& - 	& 9.00(4) 	& 8.976(46)\\
12 	& 26.106 	& 26.31(10) 	& 26.30(10) 	& 26.30(2)	& 6.1785 	& - 	& 6.15(3) 	& 6.139(35)\\
13 	& 18.281 	& 18.40(11) 	& 18.42(8) 	& 18.42(1)	& 4.4106 	& - 	& 4.39(2) 	& 4.382(27)\\
14 	& 13.297 	& 13.41(12) 	& 13.40(7) 	& 13.40(1)	& 3.258 	& - 	& 3.25(2) 	& 3.237(21)\\
15 	& 9.9725 	& 10.1(1)\tm[4] 	& 10.05(5) 	& 10.05(1)	& 2.4745 	& - 	& 2.47(1) 	& 2.458(16)\\
16 	& 7.6705 	& 7.73(5)\tm[4] 	& 7.73(5) 	& 7.728(9)	& 1.9234 	& - 	& 1.92(1) 	& 1.911(13)\\
17 	& 6.0261 	& 6.06(10)\tm[4] 	& 6.07(4) 	& 6.072(8)	& 1.5246 	& - 	& 1.52(1) 	& 1.515(10)\\
\end{tabular}
\end{ruledtabular}
\tablenotetext[1]{Reference~\cite{yang16a}}
\tablenotetext[2]{Reference~\cite{fendel07a}}
\tablenotetext[3]{Reference~\cite{stalnaker10a}}
\tablenotetext[4]{Reference~\cite{farley77a}}
\tablenotetext[5]{Reference~\cite{gerginov06a}}
\tablenotetext[6]{Reference~\cite{williams18a}}
 \end{table*}

 \begin{table}[bt]
   \caption{Fr hyperfine constants $A$ (in MHz) given by experiment and theory, and predictions using the ratio method.
Calculations were carried out for $^{211}$Fr. 
Measured data for higher states are available for $^{210,212}$Fr. We use the ratio method to predict $A$ values for the higher states of these isotopes only.
Columns are for different isotopes and should not be compared directly.
Nuclear and QED contributions cancel in the ratio (\ref{eq:ratio}), so the results are independent of which isotope is used for the theory part.
 }
\label{tab:FrRatio}
\begin{ruledtabular}
 \begin{tabular}{ccccccc}
 &  \multicolumn{2}{c}{$^{211}$Fr}& \multicolumn{2}{c}{$^{210}$Fr}	&  \multicolumn{2}{c}{$^{212}$Fr}\\
$n$	&  \multicolumn{2}{c}{Theory}& Exp	& Ratio	& Exp	& Ratio\\
\cline{2-3}\cline{4-5}\cline{6-7}
 &  $ns$&$np_{1/2}$ &\multicolumn{2}{c}{$ns$}	& \multicolumn{2}{c}{$np_{1/2}$}\\
7	& 8886.4	& 1185.5	& 7195.1(4)\tm[1]	& 	& 1192.0(2)\tm[2]	& \\
8	& 1930.8	& 367.9	& 1577.8(11)\tm[3]	& 1574	& 373.0(1)\tm[1]	& \\
9	& 763.5	& 164.7	& 622.3(3)\tm[4]	& 624	& 	& 167\\
10	& 380.4	& 88.08	& 	& 310	& 	& 89.3\\
11	& 217.2	& 52.63	& 	& 177	& 	& 53.4\\
12	& 135.6	& 33.96	& 	& 111	& 	& 34.4\\
 \end{tabular}
\end{ruledtabular}
\tablenotetext[1]{Reference~\cite{Sansonetti07a}}
  \tablenotetext[2]{Reference~\cite{grossman99a}} 
\tablenotetext[3]{Reference~\cite{simsarian99a}}
\tablenotetext[4]{Reference~\cite{gomez08a}} 
\end{table}

We note that investigating the measured hyperfine constants in terms of this trend allows us to identify instances where the midpoint of an experimental value deviates significantly from the observed trend.
For example, in Fig.~\ref{fig:rel-corr-s} it may be seen that $F^\Sigma$ derived from the Rb $9s$ measurement lies above the trend in the Rb $s$ states (this measurement also has a relatively large uncertainty compared to the other measurements).
Using the fit method, we predict the hyperfine constant to be $A_{9s} = 90.38(36)$\,MHz, in comparison to the experimental measurement of  $A^{\rm exp}_{9s} =90.9(8)$\,MHz. 
Similarly, the experimental $F^\Sigma$ for Cs $10s$ and $15s$ also lie above the trend in Fig.~\ref{fig:rel-corr-s}. 
Here, we predict $A_{10s} = 62.99(24)$\,MHz and $A_{15s} = 10.05(5)$\,MHz compared to the experimental data $A^{\rm exp}_{10s} = 63.2(3)$\,MHz and $A^{\rm exp}_{15s} = 10.1(1)$\,MHz.

\subsection{Ratio method}

We also apply the ratio method to make highly accurate predictions of the hyperfine constants for Rb and Cs, presented in Tables~\ref{tab:fittedRb} and \ref{tab:fittedCs}. 
Using the ratio method~\cite{ginges18a}, a value for the hyperfine constant $A_n$ for a state $n$ can be expressed as
\begin{equation}\label{eq:ratio}
A_n = A^{\rm th}_n \, \left({A^{\rm exp}_m}/{A^{\rm th}_m} \right),
\end{equation}
[see Eq.~(\ref{eq:ABWQED})] where $A^{\rm exp}_m$ is a measured hyperfine constant for some other state of the same angular momentum (typically taken as an excited state).
Note that the calculated $A$ values can be expressed as
$ A^{\rm th}_n = A_n(1 + \delta_n), $
where $A_n$ is the ``exact'' hyperfine constant, and $\delta_n$ is the relative deviation.
The ratio method may be used to either isolate the uncertainty for one state, when the other state can be modelled significantly better, or it may be used to make high-accuracy predictions when $\delta_n$ are comparable in magnitude and of the same sign (or small). 
When correlations are taken into account as described above, we find the value for $\delta_n$ to be the same for all $n$ to a very good approximation, and therefore cancels in the ratio (\ref{eq:ratio}); this is particularly true for the excited states, where the relative correlation corrections are smaller.
The ratio method thereby leads to very accurate predictions for hyperfine constants, so long as correlation effects are sufficiently taken into account, and there is at least one experimental value available of high accuracy ~\cite{ginges18a}.
Since the ratio  method works best when projecting from the excited states (as opposed to the ground state), we use the measurements from the lowest excited states that have the smallest uncertainties to make predictions for the higher states.
For example, for Cs, we use both the $A_{7s}^{\rm exp}$ and $A_{8s}^{\rm exp}$ experimental results to perform the predictions for the $n\geq9s$ states.
As a consistency check, we also use the experimental hyperfine constant $A_{7s}^{\rm exp}$ to predict $A_{8s}$, and  $A_{8s}^{\rm exp}$ to predict $A_{7s}$.

The leading source of uncertainty in the ratio method comes from errors in the inexact cancellation of the $\delta_n$ factors.
To this end, we calculate $\delta_n$ corresponding to our calculated $A$ values for each state using the available experimental values.
The variance in the observed $\delta_n$ values is used to estimate the uncertainty in the predicted $A$ values.
Typically, the uncertainty in the resulting $A$ values is better than $0.1\%$.
The uncertainty increases for the more highly excited states, where the exact cancellation of the $\delta_n$ terms is less certain.

The resulting predicted values up to $n=17$ are presented for Rb and Cs in Tables~\ref{tab:fittedRb} and \ref{tab:fittedCs}, respectively.
We make predictions for states where experimental data is available as a test for the method.
Note that the agreement with experiment in these cases is excellent, better than $0.05\%$ for most states.
We also present results for the first few excited states of Fr in Table~\ref{tab:FrRatio}.
With less experimental data, the uncertainty is more difficult to control, however, we expect these predictions to be accurate to the $\approx$\,0.5\% level.

\section{Summary and conclusions}
\label{sec:conclusion}

Accurate knowledge of hyperfine constants for excited states is important for extracting nuclear properties, such as nuclear magnetic moments. As atomic theory precision increases, this will extend to the Bohr-Weisskopf effect and radiative quantum electrodynamics effects that become sizeable in the strong electric field near the nucleus.
Thereby, such investigations play a significant role for testing nuclear physics models, and models for including radiative QED effects into atomic structure calculations. 
Further, the comparison of high-precision atomic structure calculations for hyperfine constants with measured values gives an important handle for understanding the accuracy of calculated wavefunctions on very small distance scales.
This is particularly important, for example, in studies of atomic parity violation.

With these motivations, we have investigated the trends in the correlation corrections for the hyperfine constants across principal quantum number $n$. We have shown that these corrections tend towards constant, non-zero values for high states. Our calculations were performed for Rb, Cs, and Fr for the ground states up to $n=18$ for $s$ and $p_{1/2}$ states. 
Our results are supported by measured values for Cs and Rb, and we have demonstrated that the smooth dependence of the relative correlation corrections on $n$ allows one to make highly-accurate predictions for the hyperfine constants. 
We have used two methods to make these predictions -- a least-squares fit to measured values, and the ratio method -- and have obtained values for the hyperfine constants for excited states of Rb and Cs with uncertainties of about 0.1\% or better.

\section{Acknowledgments}
We are grateful to M. Kozlov and V. Dzuba for useful discussions. This work was supported by the Australian Government through an Australian Research Council Future Fellowship, Project No. FT170100452. 

\bibliography{papers}

\end{document}